\documentstyle{europhys}

\def\And{{\rm and\ }}

\def\drm{{\rm d}}

\def\dif#1#2{\frac{\drm#1}{\drm#2}}

\def\stars{\bigskip\centerline{***}\medskip}

\newif\ifboo \boofalse

\begin{document}
\euro{}{}{}{}
\Date{}
\shorttitle{D. H. COBDEN {\it et al.}; FIELD EMISSION FROM A 2DEG}
\title{Current flow past an etched barrier:\\field emission from a 
two-dimensional electron gas}
\author{D. H. Cobden\inst{1}, G. Pilling\inst{1}, 
R. Parthasarathy\inst{1}, P. L. McEuen\inst{1}, 
I. M. Castleton\inst{2}, E. H. Linfield\inst{2}, 
D. A. Ritchie\inst{2}, \And G. A. C. Jones\inst{2}}
\institute{
     \inst{1} Department of Physics, University of California and 
                 Materials Sciences Division, \\Lawrence Berkeley
                 National Laboratory, Berkeley, CA 94720, USA\\
     \inst{2} Cavendish Laboratory, Madingley Road, Cambridge, 
                 CB3 OHE, UK\\
           }
\rec{}{}
\pacs{
\Pacs{73}{40Kp}{III-V semiconductor-to-semiconductor contacts, p-n
junctions, and heterojunctions}
\Pacs{79}{70.$+$q}{Field emission, ionization, evaporation, and 
desorption}
      }
\maketitle
\begin{abstract}
We find that, under appropriate conditions, electrons can pass a 
barrier etched across a two dimensional electron gas (2DEG) by field 
emission from the GaAs/AlGaAs heterojunction into a second, low-density 
2DEG formed deep in the substrate.  The current-voltage characteristics 
exhibit a rapid increase in the current at the field emission threshold 
and intrinsic bistability above this threshold, consistent with a 
heating instability occurring in the second 2DEG.  These results may 
explain similar behaviour recently seen in a number of front-gated 
devices by several groups.
\end{abstract}
Extensive studies of two-dimensional electron systems over the last two 
decades have relied on the ability to trap electrons at an interface 
between two materials with different band energies.  Most popular 
amongst these systems is the high-mobility GaAs/Al$_x$Ga$_{1-x}$As 
heterojunction \cite{Beenakker}.  Here the electrons are confined 
principally by attraction to remote donors placed in the AlGaAs.  
However, in order for there to be complete confinement there must also 
exist an electric field in the GaAs substrate pushing the electrons 
towards the heterojunction. In practice, this substrate field is 
typically provided by negative charge on residual acceptors in the 
substrate and surface boundary conditions \cite{Michels}.  A very 
interesting situation arises if this confinement field vanishes.  
The electrons are then predicted to be very weakly bound at the 
interface, by the work function of the 2DEG, as discussed by Groshev 
and Schoen \cite{Groshev}.  The work function depends sensitively on 
the density of the 2DEG and is strongly influenced by many-body 
effects such as the image potential.  Because of the weak confinement, 
field emission of electrons into the substrate may be anticipated in 
a weak electric field applied perpendicular to the interface.

In this letter we discuss a simple device that demonstrates the field 
emission of electrons from a 2DEG.  We find that a current can flow 
past an etched barrier in a 2DEG if the substrate field is adjusted 
using a positive voltage applied to a back gate.  The electrons are 
field emitted from the 2DEG and travel underneath the etched barrier.  
The magnetic field dependence demonstrates that the current is carried 
via a low-density 2DEG formed at an upside-down heterojunction below 
the uppermost GaAs substrate layer.  The observed current-voltage 
characteristics in this regime are highly nonlinear.  Above a 
threshold bias the current rises rapidly and exhibits bistability.  
We show that this bistability results from thermal runaway due to 
heating of the (initially localized) electrons in lower 2DEG by the 
field-emitted electrons from the upper 2DEG.

These results are important for three reasons.  First, they demonstrate 
a simple geometry in which the process of field emission from a 2DEG 
can be studied.  Second, they illustrate the existence of a thermal 
instability in an initially insulating, low-density 2DEG. Finally, they 
offer insight into a number of recent experiments on transport across a 
lateral barrier \cite{Pilling,Berven,Smith96,Smith97} on a GaAs/AlGaAs 
heterostructure.  The latter experiments all revealed very similar 
behaviour to that reported here.  Various explanations were put forward, 
including the heating of an accidental puddle of electrons within the 
barrier \cite{Berven}, impurities exchanging electrons \cite{Smith96}, 
and interplay with gate leakage current \cite{Duruoz}.  In many of these 
cases, however, the bistability occured under conditions when the 
substrate field vanished (after illumination), indicating that the 
mechanism proposed here may explain these experiments as well.

The device geometry, heterostructure composition and measurement 
configuration are indicated in fig. 1(a).  The back gate is a $50$ nm 
layer of n+ GaAs which is contacted separately from the 2DEG using {\it 
in situ} ion-beam patterning (see ref. \cite{Linfield}).  Above this is 
a $500$ nm AlGaAs barrier, followed by $500$ nm of GaAs at the top of 
which is the normal heterojunction.  At $V_g = 0$, a 2DEG of density 
$2.5 \times 10^{11}$ cm$^{-2}$ and mobility $3.0 \times 10^5$ 
cm$^2$V$^{-1}$s$^{-1}$ resides at this heterojunction.  The barriers in 
the 2DEG are produced by electron-beam lithography and wet etching on the 
arms of a Hall-bar mesa.  The etch width ($\sim 200$ nm) and depth 
($\sim 50$ nm) create a potential barrier at the heterojunction that 
is known to be hundreds of millivolts high \cite{Peck}; enough to make 
the barriers completely insulating at both room and low temperature 
\cite{Note1}.

Fig. 1(b) shows the $I$-$V$ characteristics of a barrier at temperature 
$T = 4.2$ K for different gate voltages $V_g$.  A bias $V$ of up to 
$100$ mV is applied to one contact, and the current $I$ is measured with 
a virtual-earth current preamplifier attached to the other contact.  
As $V_g$ is increased from zero, $I$ is zero until $V_g = 1.53$ V.  
Then, up to $V_g \gg 1.60$ V, $I$ remains zero at low $V$ but grows 
rapidly above a threshold bias.  Near this threshold there is 
bistability between a low- and a high-current state, causing 
hysteresis  between up and down sweep directions (indicated by arrows 
at $V_g = 1.54$ V) \cite{Note2}.  The threshold bias (the value of $V$ 
at which $I$ in the high-current state extrapolates to zero) decreases 
towards a limiting value $V_{lim} \approx 40$ mV as $V_g$ is increased.  
At $V_g \gg 1.60$ V the bistability disappears and simultaneously the 
conductance at $V = 0$ becomes finite.  Measurements on a number of 
barriers revealed very similar characteristics.

To understand these results, we first consider the effect of $V_g$ on 
the unetched 2DEG.  The solid line in fig. 2(a) is the density, $n_1$, 
of the 2DEG deduced from the low-field Hall coefficient of an unetched 
region, while the filled circles here are the values of $n_1$ obtained from 
Shubnikov-de Haas oscillations.  We see that $n_1$ increases linearly 
with $V_g$ up to about $1.4$ V, above which it levels off at $3.5 \times 
10^{11}$ cm$^{-2}$.  The reason for this levelling off \cite{Linfield} is 
the population of a second 2DEG, as is illustrated by the band diagrams 
in fig. 3.  At $V_g = 0$ the lower heterojunction, at the bottom of the 
$500$ nm GaAs layer, is far above $E_F$ and the electrons are tightly 
confined by the electric field in the GaAs to the upper heterojunction.  
The linear increase in $n_1$ for $V_g < 1.4$ V is determined by the 
capacitance between the upper heterojunction and the back gate, whose 
separation is $1020$ nm.  At $V_g = 1.4$ V the lower heterojunction 
reaches the Fermi level $E_F$ and a second 2DEG forms there.   For 
$V_g > 1.4$ V, the band in the GaAs remains almost flat, and $n_1$ 
remains constant. The density $n_2$ in the lower 2DEG then increases 
according to the dashed line.

Now we examine the properties of the barrier over the same range of 
$V_g$.  Fig. 2(b) shows the differential conductance, $\dif{I}{V}$, 
at $V = 0$ and $100$ mV.  Both become finite only for $V_g \geq 1.5$ V.  
The inset shows the variation with magnetic field $B$ of $\dif{V}{I} 
\equiv (\dif{I}{V})^{-1}$, at $V = 100$ mV and $V_g = 1.72$ V.  It is 
roughly linear for $B \geq 0.5$ T, suggestive of a Hall resistance.  
The same is true at other values of $V_g$.  If at each $V_g$ we 
interpret the slope as a Hall coefficient, 
$R_H = \dif{(\dif{I}{V})}{B}$, and convert it to a number density 
$(eR_H)^{-1}$, we obtain the open circles plotted in fig. 2 (a).  
The results closely follow the predicted behaviour of $n_2$.

Armed with our understanding of the behaviour of the upper and lower 
2DEGs, we can readily interpret these results.  The explanation is 
sketched in fig. 4(a).  The repulsive potential created by the etched 
surface is smallest at the lower heterojunction, as indicated by the 
contours.  It is therefore clear why the barrier only conducts once the 
gate voltage is such that the lower heterojunction is populated.  The 
current flows past the barrier along the lower 2DEG.  The onset of 
nonlinear conduction in the $I$-$V$ curves can also be understood.  To 
get from the upper to the lower 2DEG, the electrons must be field 
emitted from the upper 2DEG.  By analogy with field emission in metals, 
this happens above a characteristic electric field and hence a well 
defined source-drain bias.

To understand this in more detail, we need to consider the transfer 
rate from the upper to the lower 2DEG and the conductance $G_2$ of the 
lower 2DEG.  We first discuss the variation of $G_2$ with $V_g$.  For 
$V_g < 1.4$ V, $n_2 = 0$ and hence $G_2 = 0$.  For a range of $V_g$ 
above this, $n_2$ is low enough that localization by disorder causes 
$G_2$ to be activated, ie, $G_2 \approx G_0 \exp[-E_a / k_B T]$.  As 
$n_2$ increases, the high-$T$ conductance $G_0$ should increase while 
the activation energy $E_a$ decreases until at some point $G_2$ becomes 
measurable at $4.2$ K.  It is reasonable that this happens at 
$V_g = 1.6$ V, when $n_2 \sim 3 \times 10^{10}$ cm$^{-2}$.  In support 
of this we find from additional measurements that the linear-response 
barrier conductance is activated, with $E_a = 1.3$ mV and $G_0 = 38$ mS 
at $V_g = 1.66$ V, and that $E_a$ decreases while $G_0$ increases with 
increasing $V_g$.  This scenario is confirmed by the observation that 
the magnetoresistance of the $I$-$V$ curves accurately reflects the 
expected Hall effect of the lower 2DEG, whose two-terminal 
magnetoresistance is dominated by its Hall resistance at high $B$.  
In other words, the on-state resistance is dominated by the resistance 
of the lower 2DEG.

Having understood the role of the lower 2DEG, we can now ask how field 
emission from the upper 2DEG is related to the nonlinearity and 
bistability seen in fig. 1(b).  As $V$ is increased, the electric 
field between the upper and lower heterojunctions grows. When the 
field perpendicular to the 2DEG reaches some value we expect the rate 
of electron escape from the upper 2DEG on the negative (left) side of 
the barrier to rise rapidly, by analogy with field emission from a 
metal cathode.  This is indicated in fig. 4(b).  Variational 
calculations of the 2D subband wavefunctions imply that a 2DEG 
eventually becomes unstable (to field emission) as the substrate 
potential is lowered \cite{Hamilton}.  We can check that within our 
self-consistent simulation the onset of field emission is sudden.  We 
take the situation in fig. 3 with the lower 2DEG occupied, and 
incorporate a potential drop $\Delta V$ across the GaAs well 
\cite{Note3}.  At $\Delta V = 0$, all the lower-energy wavefunctions 
are strongly localized to one or the other side of the well.  However, 
as $\Delta V$ is increased, at some point the second subband on the 
left side of the well develops a tail on the right side.  The 
integrated probability in the tail can be approximated by 
$\exp[(\Delta V - V_{fe})/\delta]$, where $V_{fe}$ and $d$ are constants.  
Since once this tail forms electrons in the second subband can spill 
out of the upper 2DEG across the well, we identify $V_{fe}$ with the 
field emission threshold within the model.  Correspondingly, $\delta$ is 
the characteristic bias range over which field emission starts.  The 
value of $V_{fe}$ obtained depends on the acceptor concentration in the 
GaAs layer.  A concentration of $2.0 \times 10^{11}$ cm$^{-3}$ gives 
$V_{fe} = 31$ mV and $\delta = 1.6$ mV.  For all acceptor concentrations we 
find $\delta \ll V_0$, implying that the onset of field emission is indeed 
sudden.

If field emission only increased the transfer rate across the well, 
then the current would quickly be limited by $G_2$ and the $I$-$V$ 
curves would simply show a turn-on above a threshold voltage.  However, 
each field-emitted electron delivers an excess energy of up to $e \Delta V$ 
to the lower 2DEG, where $\Delta V$ is the potential difference between 
the upper and lower 2DEGs in the field emission region.  This may be 
expected to cause the electron temperature $T^*$ in the lower 2DEG to 
increase, thereby decreasing $G_2$ and increasing the current.  Within 
such a scenario a bistability arises naturally \cite{Shaw}.  We 
illustrate this by representing the barrier by the simplified 
equivalent circuit in the inset to fig. 4(c), consisting of a diode 
of turn-on voltage $V_{fe}$ in series with the activated conductance 
$G_0 \exp[-E_a/k_B T^*]$.  $T^*$ increases monotonically with $V I$, 
the power dissipated.  Our justification for neglecting the resistance 
for current returning from the lower to the upper 2DEG is that the hot 
electrons can easily traverse the well once they have passed the 
barrier.  Such a circuit exhibits an S-shaped bistable region in 
its $I$-$V$ characteristic, due to thermal runaway in the lower 
2DEG \cite{Shaw}.  Fig. 4(c) shows a characteristic generated using 
a simple proportional relationship, $T^* = \alpha I V$ and reasonable 
values of the parameters (see figure caption) chosen for similarity to 
the data at $V_g = 1.55$ V in fig. 1(b).

According to this model, the limiting threshold bias, $V_{lim}$, 
defined in the discussion of fig. 1(b), is a measure of the 
field-emission threshold, $V_{fe}$.  One may therefore estimate the 
electric field for field emission to be $E_{fe} \sim V_{lim}/d$, where 
$d$ is the GaAs well width.  For this device we get 
($40$ mV)/($0.5$ mm) $= 8 \times 10^4$ Vm$^{-1}$.  This is much lower 
than for 3D metals, where field emission typically occurs at around 
$10^9$ Vm$^{-1}$ \cite{Gomer}.  Of particular interest is the situation 
where there is no doping in the substrate, and the intrinsic work 
function \cite{Groshev} may be investigated.  Our simulations indicate 
that $E_{fe}$ is around $5 \times 10^3$ Vm$^{-1}$ in this limit.  Note that 
the threshold bias should depend on the geometry of the emitter, with a 
sharper point resulting in a lower threshold.  Future work will address 
these issues.

Finally, we note that escape from a 2DEG can easily occur whenever the 
potential is nearly flat in the GaAs substrate.  In the present devices, 
this situation is brought about by making $V_g$ sufficiently positive.  
It is known, however, that illumination with an LED at low temperatures 
also flattens the bands in the GaAs by neutralizing acceptors deep in 
the substrate \cite{Michels}.  Indeed, we have previously seen very 
similar behaviour in etched barriers on standard heterostructures with 
no back gate, but only after illumination with red light at $4.2$ K 
\cite{Pilling}, and we have subsequently found that a voltage applied 
to the chip carrier has much the same effect on the characteristics as 
has $V_g$ in the present devices.  These standard heterostructures too 
have an upside-down heterojunction around $1 \mu$m below the normal one, 
at the top of the superlattice buffer.  In Refs. 
\cite{Berven,Smith96,Smith97} infrared illumination was also applied 
before the nonlinear and bistable behaviour was observed.  We therefore 
believe that the mechanism described here can explain the puzzling 
behaviour seen in their standard heterostructure devices as well.

In conclusion, we have demonstrated a nonlinear device that relies 
on the field emission of electrons from a 2DEG trapped at a 
heterointerface to a second 2DEG beneath it.  The device exhibits 
bistable $I$-$V$'s associated with a thermal runaway in the second 
2DEG due to heating by the field-emitted electrons.  This study opens 
the way for investigations of the work function of 2D metals 
\cite{Groshev}.  Our experiments show that electrons can escape from 
2DEGs more easily than is often appreciated, and this may help to 
explain the frequent occurrence of nonlinear/bistable behaviour in a 
variety of heterostructure devices.

\stars

We thank Marc Bockrath, Kent McCormick, Alex Hamilton and Karl Brown 
for help and discussions.  This work was supported by the Office of 
Naval Research (LBNL authors), the UK EPSRC (IMC and EHL), and Toshiba 
Cambridge Research Center (DAR).

\vspace{1.5cm}
Fig. 1 - (a) Schematic view of a device, indicating the layer structure 
and measurement configuration. (b) $I$-$V$ characteristics at a series 
of $V_g$.  In each case $V$ was swept up and down once.

\vspace{0.5cm}
Fig. 2 - (a) Areal electron densities vs $V_g$.  The density of the upper 
2DEG, $n_1$, is obtained both from Hall (solid trace) and Shubnikov-de 
Haas (filled circles) measurements.  The dashed line is the predicted 
density $n_2$ of the lower 2DEG.  (b) Differential conductance vs $V_g$ 
at $V = 1$ mV and $100$ mV.  The inset shows an example of the linear 
variation of $\dif{V}{I}$ with magnetic field at $V = 100$ mV, from which 
the density values plotted as open circles in (a) are derived.

\vspace{0.5cm}
Fig. 3. - Self-consistent band profiles in the unetched heterostructure 
at three values of $V_g$, separated by $0.5$ V offsets for clarity.  
The electron density, shaded in black, is superimposed.  Uniform negative 
acceptor densities of $2.0 \times 10^{14}$ cm$^{-3}$ in the GaAs 
and $6.5 \times 10^{15}$ cm$^{-3}$ in the AlGaAs were included, the 
latter being chosen to bring the lower heterojunction 
to $E_F$ at $V_g = 1.4$ V.

\vspace{0.5cm}
Fig. 4. - Depiction of current flow past an etched barrier.  Contours of 
potential energy are sketched in, the highest being the one closest to 
the etched surface.  (a) At low bias, only equilibrium transfer occurs 
between upper and lower 2DEGs.  (b) At higher bias (negative on the 
left), electrons can be field emitted from the upper 2DEG near the 
barrier.  (c) Characteristic of the simplified equivalent circuit 
(inset) taking $V_{fe} = 38$ mV, $G_0 = 2$ mS, $E_a = 2.5$ meV, and 
$T^* = \alpha I V$ with $\alpha = 2$ K/pW.

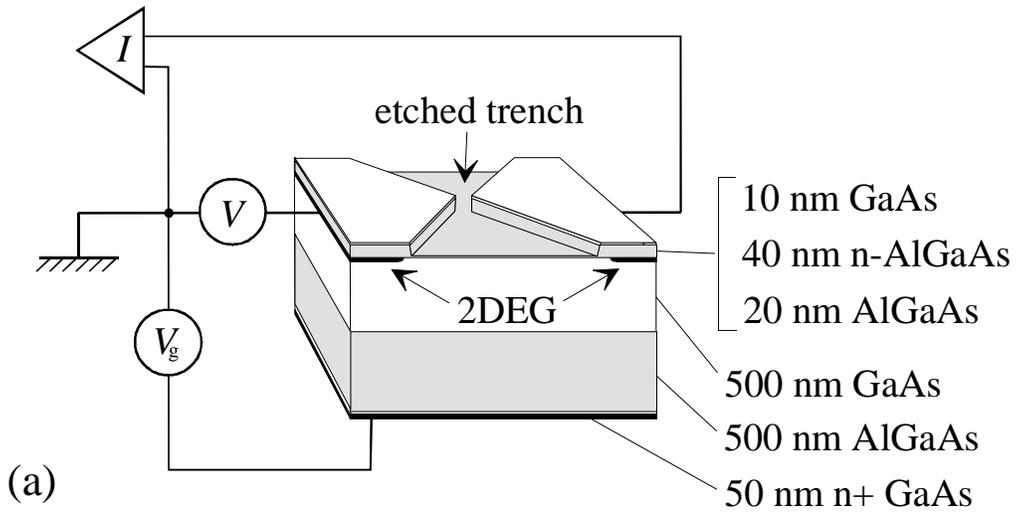
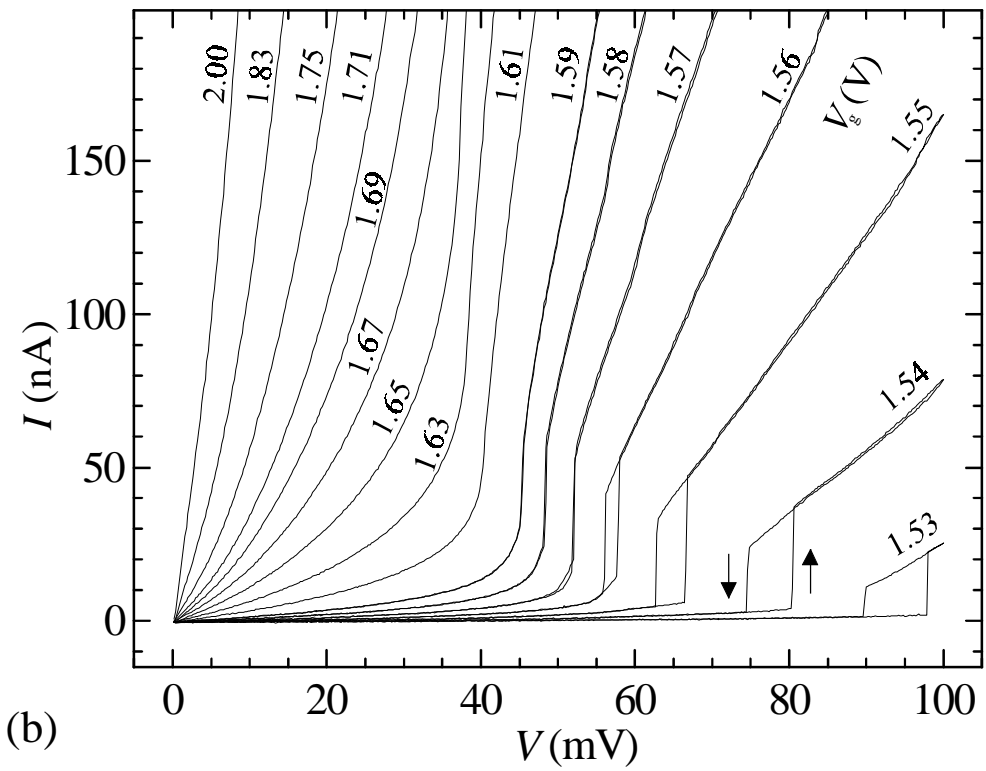
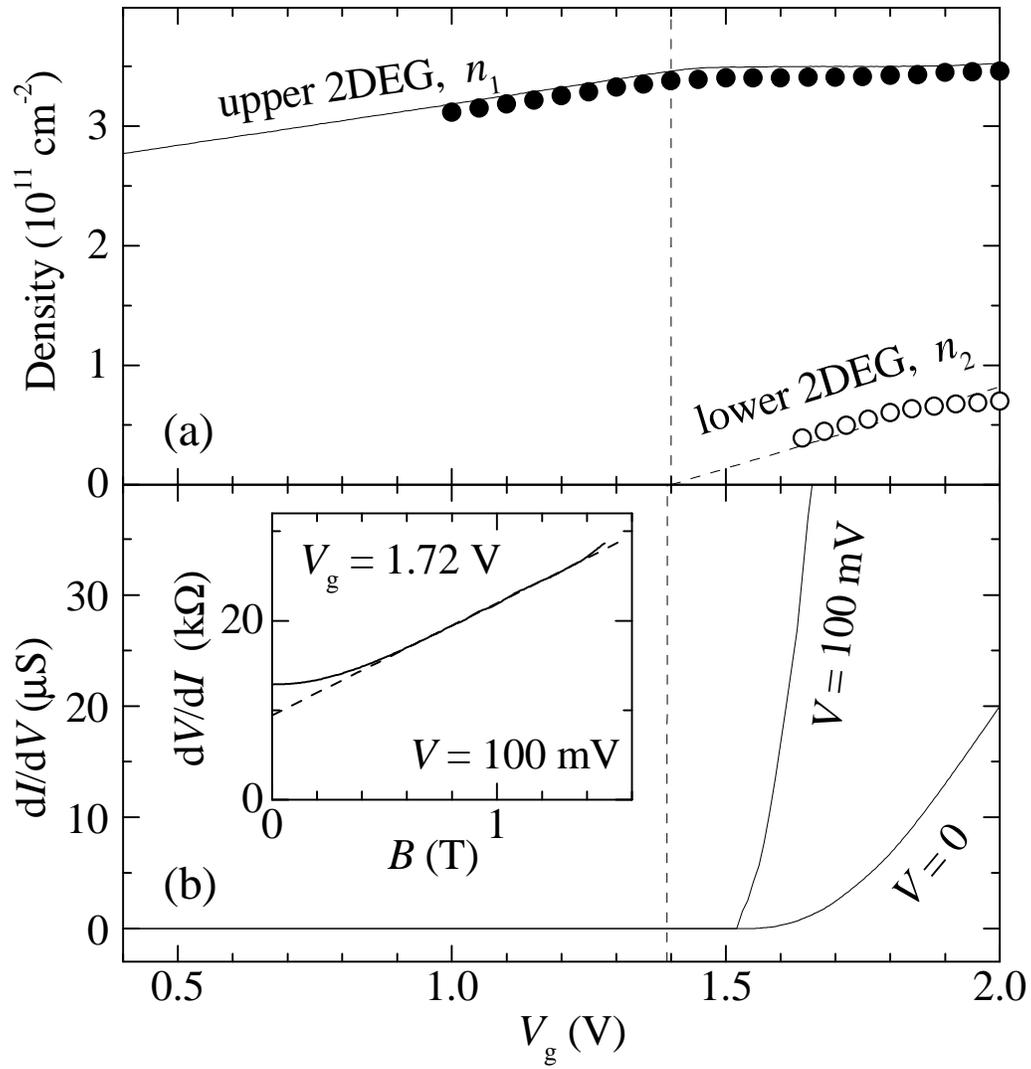
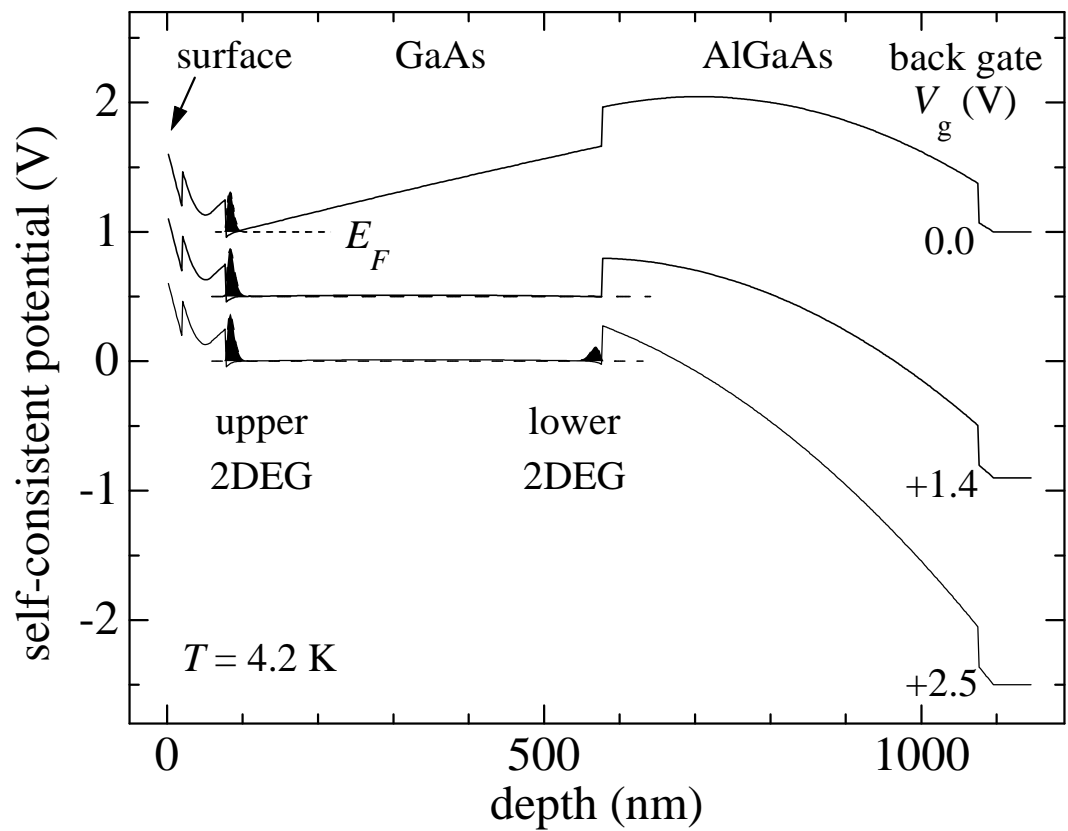
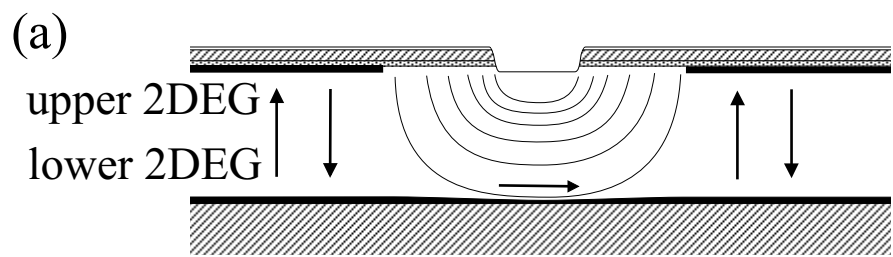
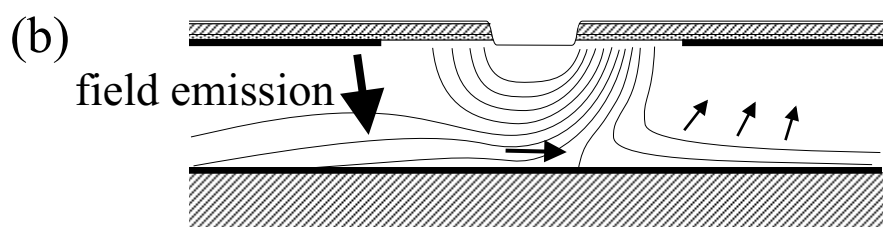
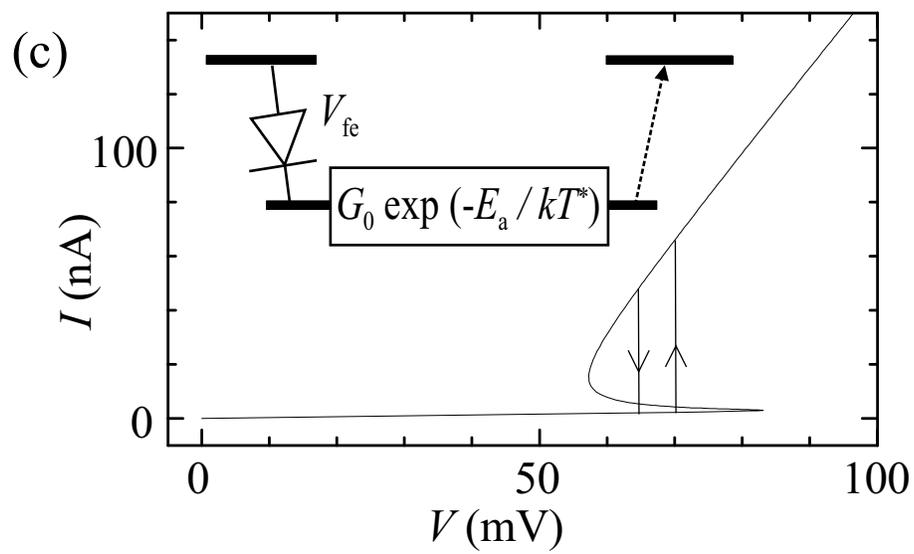
\end{document}